\documentclass[twocolumn,prb,aps,showpacs,superscriptaddress]{revtex4}
\usepackage{graphicx}

\begin{document}

\title{Evidence for Crossed Andreev Reflections in bilayers of (100) $YBa_2Cu_3O_{7-\delta}$
and the itinerant ferromagnet $SrRuO_{3}$}

\author{Itay Asulin}
\affiliation{Racah Institute of Physics, The Hebrew University of
Jerusalem, Jerusalem 91904, Israel}

\author{Ofer Yuli}
\affiliation{Racah Institute of Physics, The Hebrew University of
Jerusalem, Jerusalem 91904, Israel}

\author{Gad Koren}
\affiliation{Department of Physics, Technion - Israel Institute of
Technology, Haifa 32000, Israel}

\author{Oded Millo}
\email{milode@vms.huji.ac.il} \affiliation{Racah Institute of
Physics, The Hebrew University of Jerusalem, Jerusalem 91904,
Israel}

\begin{abstract}
Scanning tunneling spectroscopy measurements on thin epitaxial
$SrRuO_{3}/(100)YBa_2Cu_3O_{7-\delta}$ ferromagnet/superconductor
bilayers, reveal localized regions in which the superconductor
order parameter penetrates the ferromagnet to more than 26 nm, an
order of magnitude larger than the coherence length in the
ferromagnetic layer. These regions consist of narrow ($<$ 10 nm)
and long strips, separated by at least 200 nm, consistent with the
known magnetic domain wall structure in $SrRuO_{3}$. We attributed
this behavior to Crossed Andreev Reflections, taking place in the
vicinity of the magnetic domain walls.

\end{abstract}


\maketitle

In spite of a considerable research effort in the past years, a
comprehensive understanding of the proximity effect (PE) in
superconductor (S) ferromagnet (F) heterostructures has not yet
been established. Such systems are of interest since they allow a
direct investigation of the interplay between the two competing
orders of superconductivity and magnetism.  In an N/S proximity
system, where N is a normal metal in good electrical contact with
S, superconducting correlations are induced in N over a length
scale of the normal coherence length, $\xi_{N}$, while they are
weakened in the S side over a scale of the superconducting
coherence length, $\xi_{S}$ \cite{Deutscher}. The mechanism
underlying the PE at S/N interfaces is the Andreev Reflection
(AR). Upon impinging on the interface from the N side, hole-like
quasiparticles are retro-reflected as electron-like quasiparticles
with inverse spin (maintaining phase coherence over $ \xi_{N} =
\sqrt{\hbar D/k_BT}$ where $D$ is the diffusion coefficient),
while destroying Cooper pairs in the S side. Consequently, the PE
is expected to be significantly suppressed when the N side is
replaced by a ferromagnet due to spin polarization \cite{Soulen}.
Theoretical works based on the Fulde, Ferrell, Larkin and
Ovchinnikov (FFLO) mechanism \cite{Fulde,Larkin}, predict a rapid
and non-monotonic decay of the superconducting order parameter
(OP) in F, of the form $sin(x/\xi_{F})/(x/ \xi_{F})$ in the clean
limit and $exp(-x/\xi_{F})cos(x/\xi_{F})$ in the dirty limit
(where x is the distance from the interface)
\cite{Demler,Buzdin2}. The corresponding coherence lengths in F
where the exchange energy is $E_{ex}$, are $ \xi_{F} = \hbar v_{F}
/2E_{ex}$ (clean limit) and $ \xi_{F} = \sqrt{(\hbar D/2E_{ex})}$
(dirty limit), which are typically of the order of a few nm, much
shorter than $\xi_{N}$. For certain thicknesses of the F layer a
'$\pi$-state' may appear, in which the induced OP in F reverses
its sign \cite{Demler,Buzdin2}.\\
\indent Many studies confirmed these predictions and clearly
demonstrated damped OP oscillations in F and a corresponding
dependence on the F thickness of the critical current in SFS
junctions \cite{Kontos,Ryazanov}. All of these effects occur on a
length scale of a few nm, in agreement with estimates for
$\xi_{F}$. However, other experiments show a long range PE where
the penetration depth of the induced order parameter in F is two
orders of magnitude larger than $\xi_{F}$
\cite{Giroud,Petrashov}.\\
\indent The predictions concerning the S/F proximity systems
result from the singlet pairing in S, and are independent of the
symmetry (\textit{s}-wave or \textit{d}-wave) of the order
parameter \cite{Faraii,Stefanakis}. However, the anisotropy of the
\textit{d}-wave symmetry is expected to manifest itself  in the PE
\cite{Faraii} (as in the case of Au/YBCO bilayers we have
previously studied \cite{Sharoni,Asulin}), in the phase, amplitude
and period of the oscillations in F. Here too, experiments provide
contradictory results. While Ref. 15 reports short range damped
oscillations in F, data measured on SFS Josephson junctions
indicate a long range PE, sustaining for F thickness of up to 40
nm \cite{Pena,Gausepohl,Antognazza,Domel1,Domel2}.\\
\indent One possible explanation for the long range PE is given by
the formation of a strong triplet pairing amplitude component
\cite{Bergeret,Volkov}. Alternately \cite{Heikkila,Pena},
superconducting correlations can penetrate F to a distance much
longer than $\xi_{F}$ in the vicinity of magnetic domain walls
(DW) at the S/F interface via the Crossed Andreev Reflection
(CARE) effect. This process was first discussed for two spatially
separated N/S junctions \cite{Byers}, and then for three terminal
S/F hybrids \cite{Deutscher2}. Here, a spin-polarized hole
arriving from one magnetic domain is Andreev reflected as an
electron in an adjacent domain having opposite spin polarization.
In order for CARE to occur, the width of the DW must be within a
few $\xi_{S}$ \cite{Byers,Deutscher2}. Pe\~{n}a et al. \cite{Pena}
conjectured that CARE can explain their long range PE results in
SF multilayers, in spite of the fact that the DW width in their
case is $\times 10$ larger than $\xi_{S}$. Recently, evidence for
CARE was provided by magneto-transport measurements performed on
mesoscopic S/F structures consisting of conventional and
unconventional superconductors \cite{Beckmann,Giroud2,Aronov}.
However, a microscopic local-probe observation of this phenomenon
and its effect on the density of states (DOS) in the F side of the
junctions are still lacking.\\
\indent In this study we employ scanning tunneling spectroscopy on
thin epitaxial $SrRuO_{3}/(100)YBa_2Cu_3O_{7-\delta}$  (SRO/YBCO)
F/S bilayers with various SRO thicknesses. Our measurements show
that the OP (induced superconductor-like gap structure) penetrates
the SRO to a distance larger than $10\xi_F$ but only along well
defined localized lines which correlate with the underlying
magnetic DW structure. This localized long range PE may be
accounted for by the CARE process taking place along the DWs of
the SRO at the SRO/YBCO interface.\\
\indent SRO is an itinerant ferromagnet, which is ideally suited
for studying the PE with YBCO, in particular the role of the DWs.
The lattice parameters of SRO are similar to those of YBCO
\cite{Zakharov}, and therefore they can form epitaxial
heterostructures with highly transparent interfaces, essential for
the existence of AR and the PE. The DWs in SRO are $\sim 3$ nm
wide, which is comparable to the YBCO coherence length of
$\xi_{S}\sim 2$ nm, thus allowing the CARE process to occur.
Marshall et al. \cite{Marshall} have shown that, depending on the
growth orientation on $SrTiO_{3}$ substrates, the DW spacing
varies between 200 nm and $1\, \mu m$. We estimate the value of
$\xi_{F}$ in SRO to range between 1 nm in the dirty limit and 3 nm
in the clean limit.\\
\begin{figure}
\includegraphics[width=8cm]{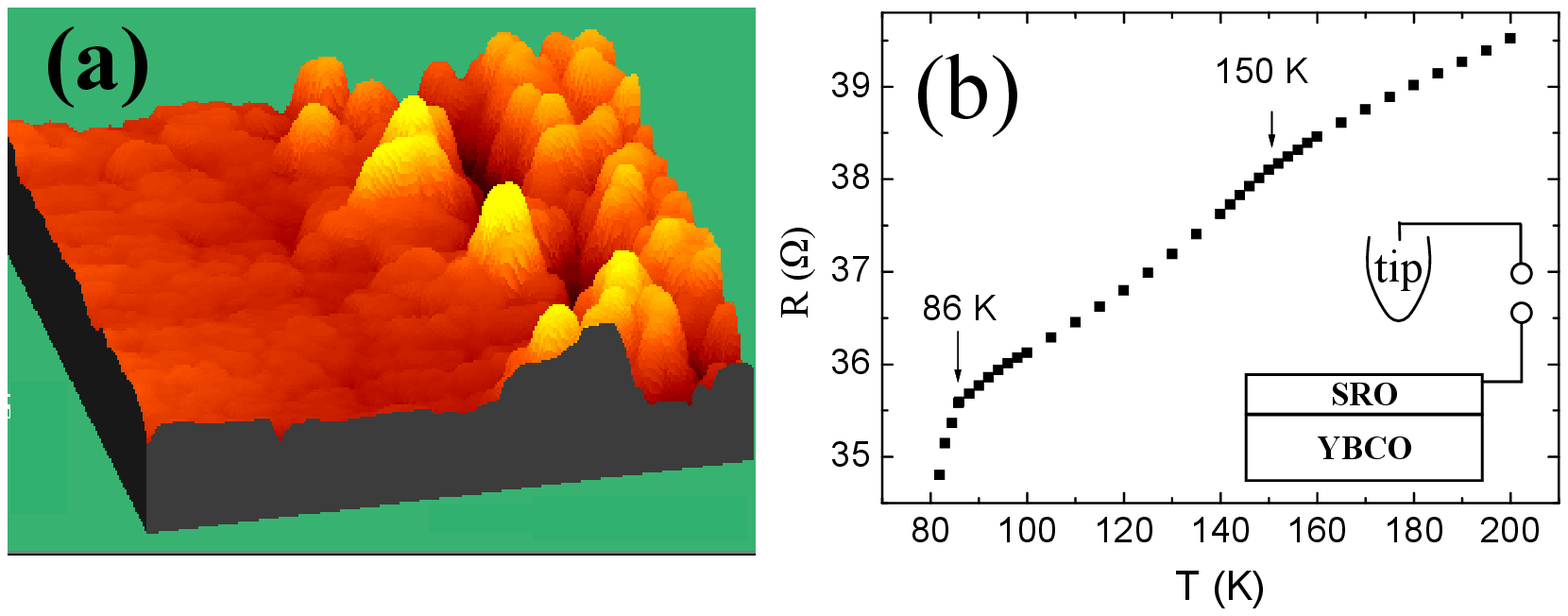}
\caption{(Color online) (a) 3D STM topographic image (0.45x0.45
$\mu m^2$) of a 13 nm thick SRO layer on a 66 nm thick (100)YBCO
film showing flat and crystalline regions typical of the bare
underlying YBCO film. (b) Resistance vs. temperature curve of a 26
nm thick SRO/ (100)YBCO bilayer showing both the ferromagnetic
transition at $\sim$ 150 K and the onset of the superconducting
transitions at 86 K. Inset - experimental setup.}\label{fig1}
\end{figure}
\indent A total of 12 bilayers of SRO (4 to 26 nm thick) on
(100)YBCO (66 nm) were prepared and measured. The (100)YBCO films
were prepared by laser ablation deposition on (100)$SrTiO_{3}$
wafers in two steps. First, a 22 nm thick template layer of YBCO
was deposited at a wafer temperature of 600 $^{0}$C. Then, a
second 44 nm thick YBCO layer was prepared at 760 $^{0}$C. This
produced films with two coexisting a-axis phases on about 95
percent of the film's area (verified by X-ray diffraction). One
phase consists of small crystallites, a few unit-cells in height
($\sim 2$ nm), while the other is composed of large areas,
atomically smooth on scales of 100 nm (see Ref. 31). The bare YBCO
films showed transition temperatures at around 88 K with a
transition width of about 2 K, implying nearly optimally doped
homogeneous films. Tunneling spectra obtained on the smooth
regions of the bare YBCO sample featured 16-18 mV gaps, mainly
U-shaped, further verifying the a-axis orientation
\cite{Sharoni2}. The SRO layer was deposited \textit{in-situ} on
the a-axis YBCO films at 800 $^{0}$C substrate temperature, under
100 mTorr of oxygen flow. The bilayer was then annealed in 50 Torr
of oxygen for 1h at 430 $^{0}$C. Morphological features
reminiscent of the flat and crystalline regions mentioned above
were apparent also on the SRO-coated films. This is shown in Fig.
1a, presenting an STM image of a 13 nm thick SRO layer overcoating
a (100)YBCO film. The samples were transferred from the growth
chamber in a dry atmosphere and introduced into our cryogenic STM
after being exposed to ambient air for less than 10 min. R(T)
curves of the SRO/(100)YBCO bilayers (for SRO layers thicker than
8 nm) clearly showed both the ferromagnetic transition at
$\sim$150 K and the superconducting transition onset around 86-90
K as seen in Fig. 1b, indicating that SRO layers at these
thicknesses are still ferromagnetic but have no
\textit{pronounced} effect on the bulk
superconductivity. (Zero resistance was obtained in Fig. 1b at 78 K).\\
\begin{figure}
\includegraphics[width=8cm]{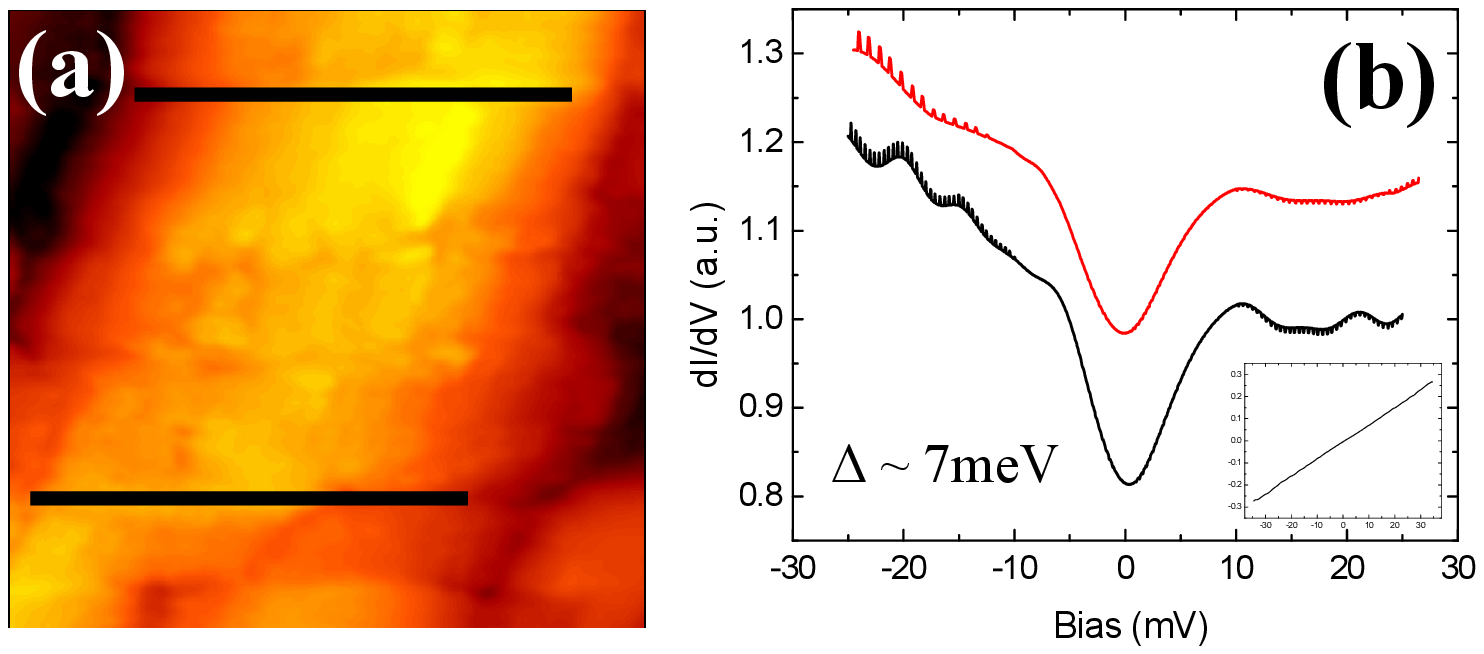}
\caption{(Color online) (a) $300\times 300$ nm$^2$ 2D STM image of
a 9 nm thick SRO/YBCO bilayer on which an induced OP was detected
along two elongated strips (marked). (b) Tunneling spectra taken,
correspondingly, at the upper and lower strips marked in (a) (the
upper curve is shifted vertically for clarity). Outside these
gapped strips, only Ohmic I-V curves were measured (inset).
}\label{fig2}
\end{figure}
\indent The STM data presented here were all acquired at 4.2 K
using a (normal metal) Pt-Ir tip as seen in the inset to Fig. 1b.
The tunneling spectra were taken at specific well-defined
locations correlated with the surface topography while momentarily
disconnecting the feedback loop. Figure 2a presents a topographic
image of a 9 nm thick (at least 3 $\xi_{F}$) SRO layer overcoating
a (100)YBCO film. Within this region we observed two parallel
strips, 200 nm apart (marked in black lines), along which gapped
tunneling spectra were measured. Such strips will hereafter be
referred to as 'gapped strips'. The dI/dV vs. V curves presented
in Fig. 2b were taken correspondingly on the lower and upper
gapped strips, both showing a pronounced mini-gap of $\Delta\sim
7$ mV and a normalized (with respect to the normal conductance)
zero bias conductance (ZBC) of $\sim 0.8$. We note that above
$T_c$, no gaps were observed. FFLO-type PE theories
\cite{Demler,Buzdin2} predict that the proximity induced OP should
virtually vanish for this SRO thickness ($\sim 3\xi_{F}$) and
consequently the ZBC should be 1. Indeed the area between the two
marked strips featured exclusively Ohmic (gapless) I-V curves
(inset to Fig. 2b). In most cases, the location of the strips had
no correlation to detectable topographic features. This excludes
the possibility that the induced gap in the DOS originates from a
proximity to grain boundaries, cracks or other defects where the
SRO thickness might be lower than the nominal value or that
ferromagnetism could be locally suppressed. The distance between
such gapped strips was in many cases (as in Fig. 2a) 200 nm, but
larger separations were also observed, consistent
with the domain structure of SRO reported in Ref. 30.\\
\indent A more thorough mapping of the spatial evolution of the
DOS in the vicinity of such a strip is presented in Fig. 3. Figure
3b depicts a $260\times 260$ nm$^2$ topographic image of an 18 nm
thick (at least 6 $\xi_{F}$) SRO layer overcoating a (100) YBCO
film, where the center of a gapped strip is marked by the broken
blue arrow. The spectra presented in Fig. 3c were acquired
sequentially at equal steps along this strip. Clearly, the gap in
the DOS is continuous and remarkably constant \textit{along} the
strip, over the whole length that was measured (about 200 nm). The
tunneling spectra depicted in Fig. 3d were taken in a similar
manner \textit{across} the strip (solid white arrow in 3b) over a
total length of 10 nm. Evidently, the width of the gapped area in
the central region of this cross-line is less than 8 nm,
comparable to the width of the DWs in SRO ($\sim3 nm$)
\cite{Marshall}. Within that narrow region, the gap is continuous
and has a fixed width of 6.8 meV, while the ZBC shows small
variations. Outside this central region, the gap decays abruptly
over a length of a nm and the DOS becomes normal. (see the
projection onto the x-y plane in Fig. 3d). We note that the gap
width and ZBC could vary from one strip to another on a specific
film, but on average the gap
structure weakened with increasing SRO layer thickness.\\
\begin{figure}
\includegraphics[width=8cm]{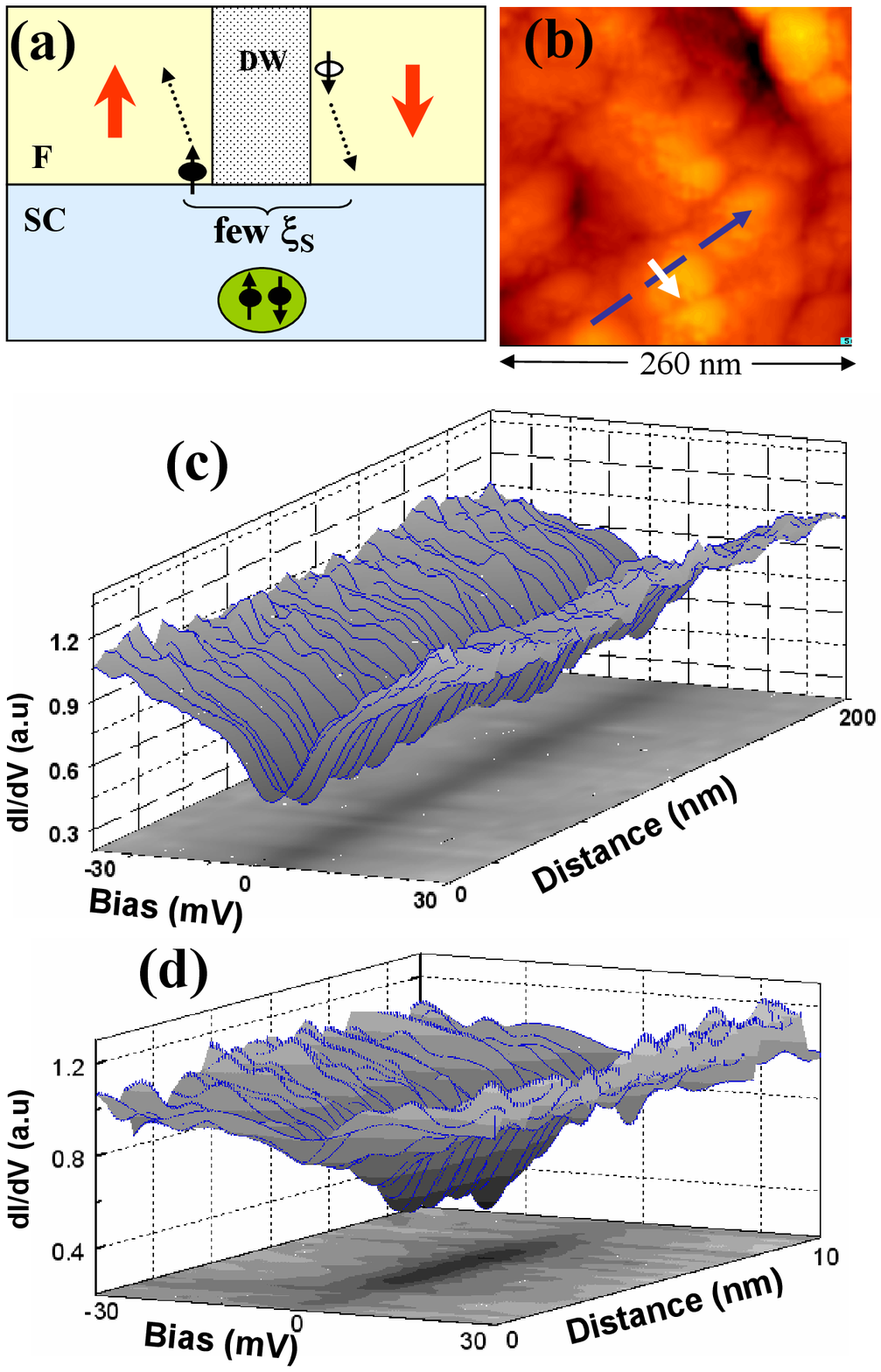}
\caption{(Color online) STM measurement demonstrating the spatial
evolution of the DOS within a 'gapped strip' in the SRO. (a) A
schematic CARE diagram. (b) $260\times260 $ nm$^2$ topographic
image of a 18 nm thick SRO/(100)YBCO bilayer. (c) Tunneling
spectra taken sequentially at fixed steps along the center of the
gapped strip (marked by the blue arrow in (b)). (d) Tunneling
spectra acquired across the gapped strip (solid white arrow in
(b)).}\label{fig3}
\end{figure}
\indent Interestingly, no gapped strips were detected on SRO films
thinner than $2\xi_{F}$. Instead, the OP seemed to penetrate the F
layer over large areas but in a non-uniform manner: for a given
SRO layer thickness, gaps with a very wide distribution of ZBC
(0.5 to 0.85) and gap width (3.8 mV to 7.5 mV) were observed. The
lower panel of Fig. 4 depicts two sets of dI/dV curves acquired at
two different areas ($\sim 100\times 100$ nm$^2$ each) on a 6 nm
thick SRO layer overcoating a (100)YBCO film. The wide
distribution of the gap features cannot be solely due to SRO
thickness variations. Possibly, in this low SRO thickness regime,
ferromagnetism might be weakened in parts of the F layer and the
domain structure may be lost. Indeed, the kink seen at 150 K in
the R(T) curve of Fig. 1b could hardly be observed on these
bilayers. The upper panel of Fig. 4 depicts a typical dI/dV curve
obtained on an area adjacent to the area where the small gap
curves below it were measured. This curve shows a peak in the DOS
at zero bias, the shape and size of which (compared to the curves
plotted below it) are consistent with the
formation of a $\pi$-state.\\
\begin{figure}
\includegraphics[width=8cm]{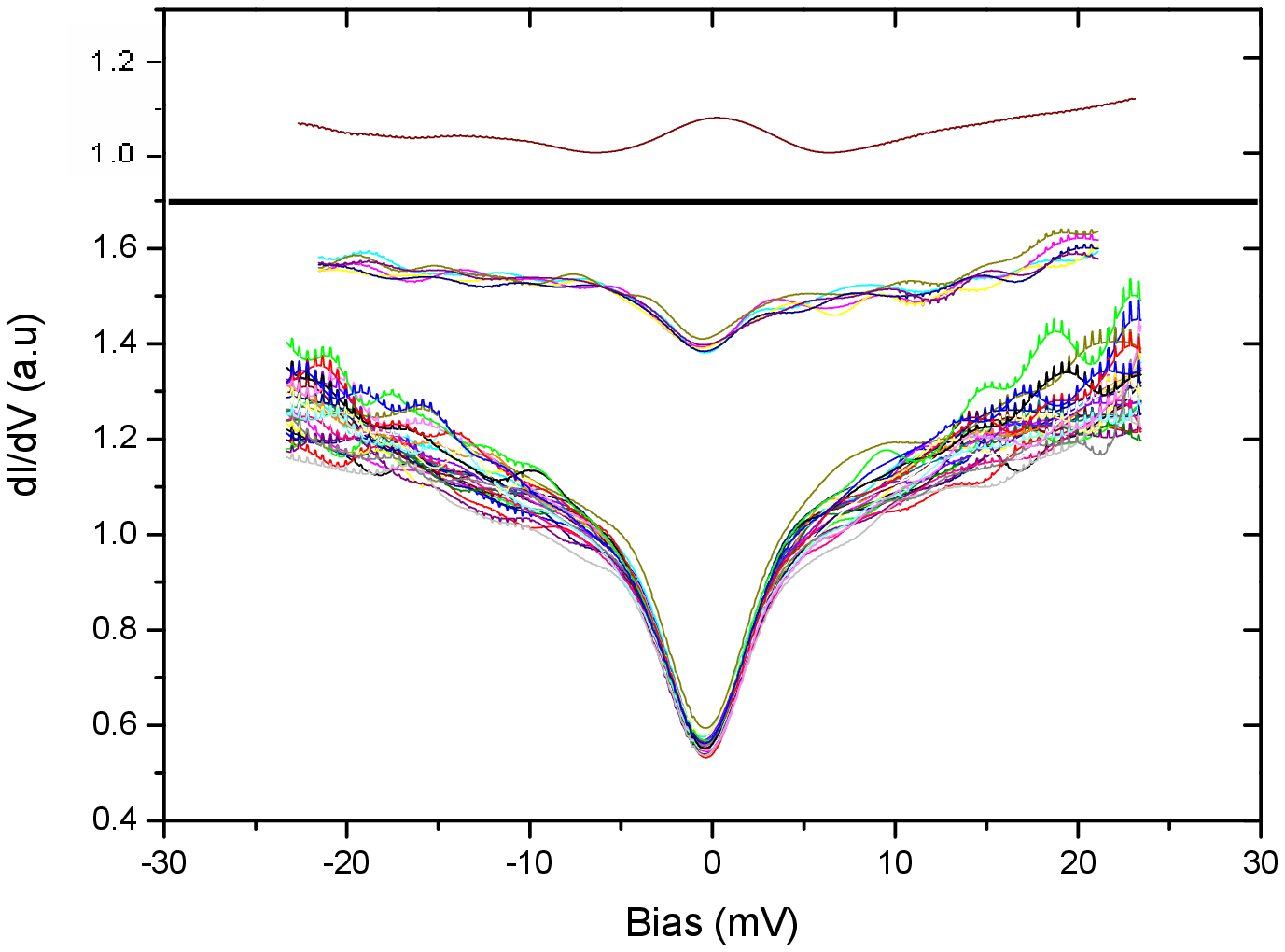}
\caption{(Color online) Lower panel: two sets of tunneling spectra
taken at different areas ($100\times 100\, nm^2$ each) on a 6 nm
thick SRO/(100)YBCO bilayer. Gap size and ZBC corresponding to the
upper and lower sets are: 3.8 mV, 0.85 and 4.5 mV, 0.56,
respectively. Upper panel: a dI/dV curve showing a peak at zero
bias, possibly manifesting a $\pi-state$.}\label{fig4}
\end{figure}
\indent We believe that the origin for the observed localized and
long-range ($\geq 10\xi_{F}$) penetration depth of the OP into the
SRO layer is the CARE process, taking place along the DWs at the
SRO/YBCO interface (see Fig. 3a). Consequently, superconducting
correlations can penetrate F quite efficiently in the vicinity of
a DW. Naively speaking, Cooper pairs are injected into the F layer
and can diffuse deeper inside only along a DW up to distances
comparable to those of the PE in the S/N case \cite{Sharoni}.
Recall that the DW width is comparable to $\xi_S$ and much smaller
than the phase coherence length ($\sim \xi_N$) at 4.2 K, thus the
conditions required for the local long range PE are satisfied. The
narrow width and the spacing of the elongated gapped strips in our
measurements are in agreement with the known configuration and
size of the DWs in SRO \cite{Marshall}. Our results support the
recent magneto-transport evidence for CARE in YBCO/SRO/YBCO
junctions with similar SRO layer thicknesses \cite{Aronov}, and
corroborate the prediction that the total Andreev conductance of
an S/F interface is proportional to the total length of the DWs
crossing it \cite{Chtchelkatchev}. The triplet pairing scenario
for the long range PE is unlikely in our case since it would have
resulted in a penetration of the OP \textit{all over} the SRO
layer. However, we cannot exclude the possibility that the long
range PE may be due to the reduced spin polarization inside the
DWs that may locally enhance the conventional AR.\\
\indent We suggest that the CARE process may also account for the
results reported by Gausepoh et. al.\cite{Gausepohl} and D\"{o}mel
et al.\cite{Domel1}. In both cases, supercurrents were observed in
YBCO/SRO/YBCO ramp junctions with 20 nm thick SRO barriers. In the
former, nonuniformity in the supercurrent density over the area of
the junctions was inferred from the magnetic response  and related
to the existence of favorable interface regions, less than 10 nm
wide, through which the supercurrent flows. In the latter, the
results were attributed to resonant tunneling through localized
states. Our data imply that the localized behavior in both cases
may possibly be an effect of the existence of DWs in the SRO layer
through which the superconducting electrodes couple. In the
opposite case, where no DWs are present at an F/S interface, the
'conventional' short range PE is expect to take place, as was
recently demonstrated experimentally \cite{Aumentado}.\\
\indent In summary, we found a remarkably localized and long
ranged PE in bilayers of SRO on (100)YBCO. For SRO layers thicker
than $\sim 3\xi_F$, the OP penetrates the SRO to a distance larger
than $10\xi_F$ only along well defined localized lines which
correlate with the underlying magnetic DW structure. This
localized long range PE is thus attributed to the CARE process
taking place along the DWs of the SRO layer.\\
\indent The authors are grateful to G. Deutscher, and L. Klein,
for useful discussions. This research was supported in part by the
Israel Science Foundation, Center of Excellence program (grant \#
1564/04).


\begin{references}

\bibitem {Deutscher}
G. {Deutscher} and P. G. {De Gennes}, {\emph{Superconductivity}}
(Marcel Dekker, Inc., New York, 1969).

\bibitem {Soulen}
R. J. {Soulen} et al., Science \textbf{282}, 85 (1998).

\bibitem {Fulde}
P. {Fulde}, and R. A. {Ferrell}, Phys. Rev. \textbf{135},A550
(1964).

\bibitem {Larkin}
A. {Larkin}, and Y. {Ovchinnikov}, Sov. Phys. JETP \textbf{20},762
(1965).

\bibitem {Demler}
E. A. {Demler}, G. B. {Arnold}, and M. R. {Beasley}, Phys. Rev. B
\textbf{55}, 15174 (1997).

\bibitem {Buzdin2}
A. {Buzdin}, Rev. Mod. Phys. \textbf{77}, 935 (2005).


\bibitem {Kontos}
T. {Kontos} et al., Phys. Rev. Lett. \textbf{86}, 304 (2001).

\bibitem {Ryazanov}
V. V. {Ryazanov} et al., Phys. Rev. Lett. \textbf{86}, 2427
(2001).


\bibitem {Giroud}
M. {Giroud} et al., Phys. Rev. B \textbf{58}, R11872 (1998).

\bibitem {Petrashov}
V. T. {Petrashov} et al., Phys. Rev. Lett. \textbf{83}, 3281
(1999).

\bibitem {Faraii}
Z. {Faraii} and M. {Zareyan}, Phys. Rev. B \textbf{69}, 014508
(2004).

\bibitem {Stefanakis}
N. {Stefanakis} and R. {M\'{e}lin}, J. Phys: Condensed Matter
\textbf{15}, 3401 (2003).

\bibitem {Sharoni}
A. {Sharoni} et al., Phys. Rev. Lett. 92, 017003 (2004).

\bibitem {Asulin}
I. {Asulin} et al., Phys. Rev. Lett. \textbf{93}, 157001 (2004).

\bibitem {Freamat}
M. {Freamat} and K. W. {Ng}, Phys. Rev. B \textbf{68}, 060507(R)
(2003).

\bibitem {Pena}
V. {Pe\~{n}a} et al., Phys. Rev. B \textbf{69}, 224502 (2004).

\bibitem {Gausepohl}
S. C. {Gausepohl}, et. al., Appl. Phys. Lett. \textbf{67}, 1313
(1995).

\bibitem {Antognazza}
L. {Antognazza} et al., Appl. Phys. Lett. \textbf{63}, 1005
(1993).

\bibitem {Domel1}
R. {D\"{o}mel} et al., Appl. Phys. Lett. \textbf{67}, 1775 (1995).

\bibitem {Domel2}
R. {D\"{o}mel} et al., Supercond. Sci. Technol. \textbf{7}, 277
(1994).

\bibitem {Bergeret}
F. S. {Bergeret}, A. F. {Volkov}, and K. B. {Efetov}, Phys. Rev.
Lett. \textbf{86}, 4096 (2001).

\bibitem {Volkov}
A. F. {Volkov}, F. S. {Bergeret}, and K. B. {Efetov}, Phys. Rev.
Lett. \textbf{90}, 117006 (2003).

\bibitem {Heikkila}
T. {Heikkil\"{a}},  Doctoral Dissertation, Helsinki University of
Technology, 2002.

\bibitem{Byers} J. M. Byers and M. E. Flatt$\rm \acute{e}$, Phys. Rev. Lett.
\textbf{74}, 306 (1995).

\bibitem {Deutscher2}
G. {Deutscher} and D. {Feinberg}, Appl. Phys. Lett. \textbf{76},
487 (2000).

\bibitem {Beckmann}
D. {Beckmann}, H. B. {Weber}, and H. v. {L\"{o}hneysen}, Phys.
Rev. Lett. \textbf{93}, 197003 (2004).

\bibitem {Giroud2}
M. {Giroud} et al., Eur. Phys. J. B \textbf{31}, 103 (2003) .

\bibitem {Aronov}
P. {Aronov}, and G. {Koren}, Phys. Rev. B. \textbf{72}, 184515
(2005).

\bibitem {Zakharov}
N. D. {Zakharov} et al., J. Mater. Res. \textbf{14}, 4385 (1999).

\bibitem {Marshall}
A. F. {Marshall} et al., J. Appl. Phys. \textbf{85}, 4131 (1999).

\bibitem {Sharoni2}
A. {Sharoni} et al., Europhys. Lett., \textbf{62}, 883 (2003)

\bibitem {Chtchelkatchev}
N. M. {Chtchelkatchev} and I. S. {Burmistrov}, Phys. Rev. B.
\textbf{68}, 140501(R) (2003).

\bibitem {Aumentado}
J. {Aumentado} and V. {Chandrasekhar}, Phys. Rev. B \textbf{64},
054505 (2001).


\end{references}
\end{document}